\documentclass[%
twocolumn,
showpacs,            
preprintnumbers,
prd
a4paper,
 superscriptaddress, 
nofootinbib,         
tightenlines,        
amsmath,amssymb
floats,floatfix      
]{revtex4-1}

\usepackage[usenames]{xcolor}
\newcommand{\red}[1]{{\bf \color{red} #1\color{black}}}

\usepackage{parskip}
\usepackage{hyperref}
\usepackage{graphicx}
\usepackage{natbib}

\usepackage{amsfonts}
\usepackage{amssymb}
\usepackage{amsmath}

\newcommand{\kmax}{$k_\mathrm{max}$}
\newcommand{\Neff}{$N_\mathrm{eff}$}
\newcommand{\mnu}{$\sum m_\nu$}
\newcommand{\Mnu}{$\sum m_\nu$}
\newcommand{\km}{\, \mathrm{km}}
\newcommand{\s}{\, \mathrm{s}}
\newcommand{\Mpc}{\, \mathrm{Mpc}}
\newcommand{\eV}{\, \mathrm{eV}}
\newcommand{\hMpc}{$\,h\,\textrm{Mpc}^{-1}$}

\newcommand{\lsim}{\mbox{$\:\stackrel{<}{_{\sim}}\:$} }

\newcommand{\Mnuthree}{$\Lambda\mathrm{CDM}_{3\nu}$}
\newcommand{\Mnutwo}{$\Lambda\mathrm{CDM}_{1+2\nu}$}
\newcommand{\Mnutwoone}{$\Lambda\mathrm{CDM}_{2+1\nu}$}
\newcommand{\Mnufour}{$\Lambda\mathrm{CDM}_{3+1\nu}$}
\newcommand{\LCDM}{$\Lambda\mathrm{CDM}$}

\newcommand\eqnref[1]{%
 Eqn.~\ref{eqn:#1}}

\newcommand\tabref[1]{%
Tab.~\ref{tab:#1}}

\newcommand\figref[1]{%
Fig.~\ref{fig:#1}}

\newcommand\secref[1]{%
Sec.~\ref{sec:#1}}

\newcommand{\jcap}{Journal of Cosmology and Astroparticle Physics}

\newcommand{\mnras}{Mon. Not. R. Astron. Soc.}
\newcommand{\pasa}{Publications of the Astronomical Society of Australia}

\begin{document}

\title{Combining Planck with Large Scale Structure gives strong neutrino mass constraint}
\author{Signe Riemer-S\o rensen}
\email{Email: signe@physics.uq.edu.au}
\affiliation{School of Mathematics and Physics, University of Queensland, Brisbane, QLD 4072, Australia}
\affiliation{Institute of Theoretical Astrophysics, University of Oslo, PO 1029 Blindern, 0315 Oslo, Norway}
\author{David Parkinson}
\affiliation{School of Mathematics and Physics, University of Queensland, Brisbane, QLD 4072, Australia}
\author{Tamara M.\ Davis}
\affiliation{School of Mathematics and Physics, University of Queensland, Brisbane, QLD 4072, Australia}

\begin{abstract}
We present the strongest current cosmological upper limit on the neutrino mass of $\sum m_\nu < 0.18 \eV$ (95\% confidence). It is obtained by adding observations of the large-scale matter power spectrum from the WiggleZ Dark Energy Survey to observations of the cosmic microwave background data from the Planck surveyor, and measurements of the baryon acoustic oscillation scale. The limit is highly sensitive to the priors and assumptions about the neutrino scenario. We explore scenarios with neutrino masses close to the upper limit (degenerate masses), neutrino masses close to the lower limit where the hierarchy plays a role, and addition of massive or massless sterile species. 
\end{abstract}

\maketitle

\section{Introduction}
The quest to determine the neutrino mass scale has been dominated by lower limits from particle physics experiments complemented by upper limits from cosmology. Recently the allowable mass window was narrowed by the Planck surveyor's measurements of the cosmic microwave background (CMB) providing an upper limit on the sum of neutrino masses\footnote{Planck+WMAP polarisation data+high-$\ell$ from the South Pole and Atacama Cosmology Telescopes} of $\sum m_\nu < 0.66 \eV$ (all quoted upper limits are 95\% confidence), or $\sum m_\nu < 0.23\eV$ when combined with baryon acoustic oscillation (BAO) measurements \cite[][]{PlanckXVI:2013}. The BAO tighten the constraint by breaking the degeneracies between other parameters (primarily the matter density and expansion rate), but do not themselves encode any significant information on the neutrino mass \cite{Hamann:2010}.

On the other hand, the full shape of the matter power spectrum of large scale structure {\em does} contain significant information on the neutrino mass. Massive neutrinos affect the way large-scale cosmological structures form by slowing the gravitational collapse of halos on scales smaller than the free-streaming length at the time the neutrinos become non-relativistic. This leads to a suppression of the small scales in the galaxy power spectrum that we observe today, and consequently we can infer an upper limit on the sum of neutrino masses \cite{Hu:1998,Lesgourgues:2006}. The shape of the matter power spectrum was not used by the Planck team to avoid the complexities of modelling non-linear growth of structure. They admit that non-linear effects may be small for $k<0.2$\hMpc, but justify their choice with ``there is very little additional information on cosmology once the BAO features are filtered from the [power]spectrum, and hence little to be gained by adding this information to {\em Planck}'' \cite{PlanckXVI:2013}.  

In this paper we show that adding matter power spectrum data to Planck+BAO data does improve the neutrino mass constraint by $0.05\eV$ to $\sum m_\nu < 0.18\eV$. 
Cosmological neutrino mass constraints now push so close to the lower limit of $\sum m_\nu > 0.05\eV$ from neutrino oscillation experiments \cite{Fukuda:1998,Beringer:2012,Forero:2012} that the ordering of the neutrino masses (hierarchy) may  play a role. In this paper we explore various hierarchy assumptions including the existence of extra relativistic species.

We only consider the matter power spectrum at large scales ($k<0.2$\hMpc) for which non-linear corrections (from structure formation and redshift space distortions combined) happen to be small for the blue emission line galaxies that we use from the WiggleZ Dark Energy Survey. These can be calibrated using simulations \cite{Parkinson:2012}. 

The paper is organised as follows:
\secref{models} describes the cosmological  scenarios we explore, while
\secref{method} gives an overview of the observational data and analysis methods.
In \secref{results} we present the results and discuss how they are affected by the various neutrino assumptions, before summarising our findings in \secref{summary}.

\section{Neutrino models} \label{sec:models}
We compute neutrino mass constraints for a number of different models corresponding to different neutrino scenarios: 
\begin{itemize}
\item neutrinos close to the upper mass limit where the masses are effectively degenerate,
\item neutrinos close to the lower mass limit where the hierarchy plays a role, and
\item the addition of massive or massless sterile species. 
\end{itemize}

For each scenario (described in more detail below) we fit the data to a standard flat $\Lambda$CDM cosmology with the following parameters: the physical baryon density ($\Omega_\mathrm{b}h^2$), the physical dark matter density ($\Omega_\mathrm{cdm}h^2$), the Hubble parameter at $z=0$ ($H_0$), the optical depth to reionisation ($\tau$), the amplitude of the primordial density fluctuations ($A_s$), and the primordial power spectrum index ($n_s$). 

In addition we vary the sum of neutrino masses, $\sum_{i=0}^{i=N_\nu} m_{\nu,i}$, where $N_\nu$ is the number of massive neutrinos. The total energy density of neutrino-like species is parametrised as $\rho_\nu = N_\mathrm{eff} T_\nu^4 7\pi^2/120$ where \Neff{} is the effective number of species $N_\mathrm{eff} = N_\nu +\Delta N$. When considering standard \LCDM{} the neutrino parameters are fixed to $\sum m_\nu = 0.06\eV$ and \Neff{} $=3.046$, where the 0.046 accounts for the increased neutrino energy densities due to the residual heating provided by the $e^+e^-$-annihilations because the neutrinos do not decouple instantaneously and the high-energy tail remains coupled to the cosmic plasma \cite{Mangano:2005,Riemer-Sorensen:2013review,Lesgourgues:2012review}.


There is no evidence from cosmological data that \LCDM{} requires a non-zero neutrino mass to provide a better fit \cite{Feeney:2013}, but the prior knowledge from particle physics justifies, and indeed requires, the inclusion of mass as an extra parameter. We know that at least two neutrinos have non-zero masses because oscillation experiments using solar, atmospheric, and reactor neutrinos have measured mass differences between the three standard model species to be $\Delta m_{32}^2 = |(2.43^{+0.12}_{-0.08}) \times10^{-3}|\eV^2$ and $\Delta m_{21}^2 = (7.50\pm0.20)\times10^{-5}\eV^2$  \cite{Fukuda:1998,Beringer:2012}. The Heidelberg-Moscow experiment has limited the mass of  the electron neutrino to be less than $0.35\eV$ (90\% confidence level) using neutrino-less double $\beta$-decay \cite{Klapdor:2006}, but does not require the neutrinos to be massive. No current experiment has sufficient sensitivity to measure the absolute neutrino mass.

\begin{figure}
	\centering
	\includegraphics[width=0.99\columnwidth]{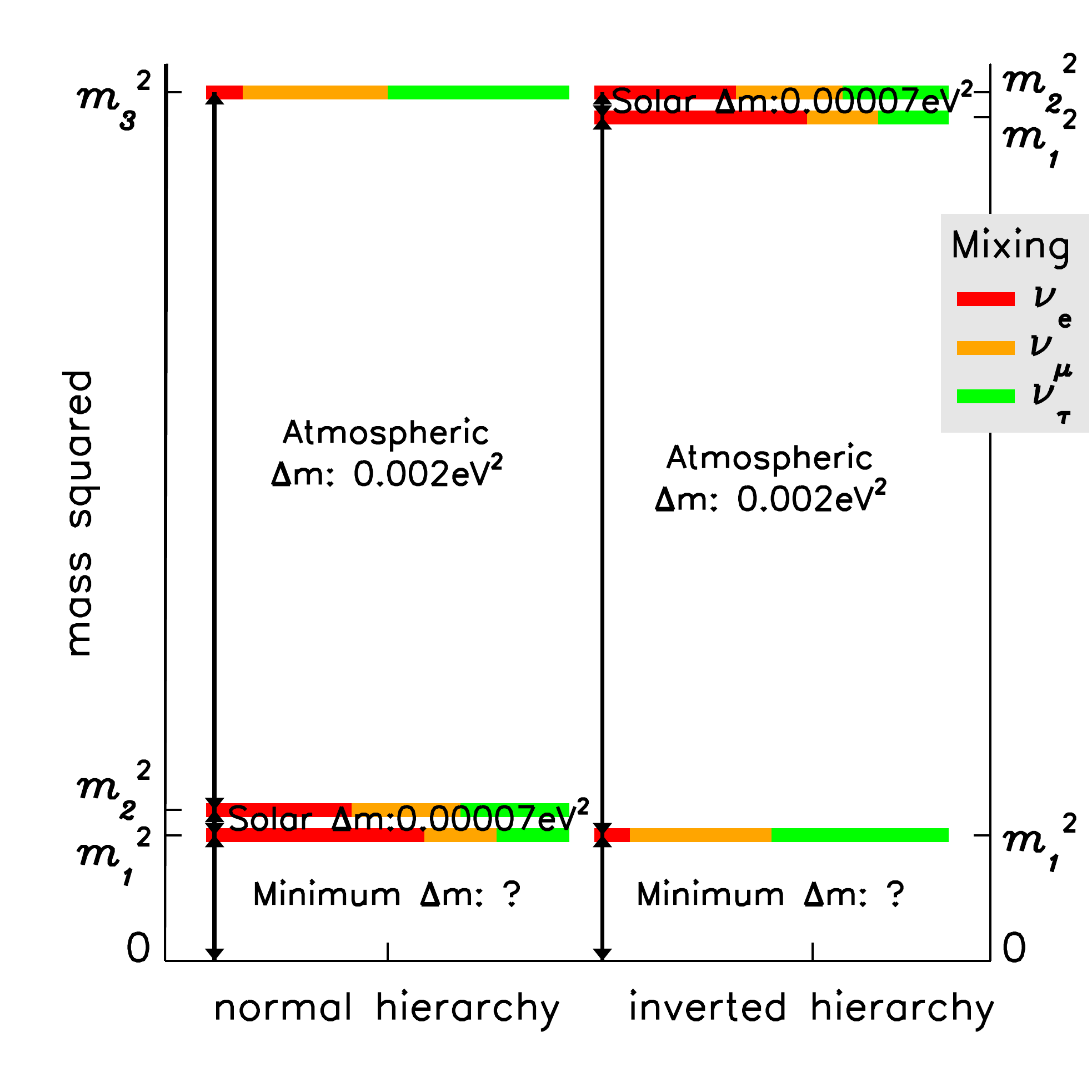}
	\caption{The current knowledge of neutrino masses and mixing between the interaction eigenstates as obtained from neutrino oscillation experiments \cite{Fukuda:1998,Beringer:2012} for the three normal/active neutrinos ($\nu_e,\nu_\mu, \nu_\tau$). If the value of $\Delta m$ is large, the mass differences are much smaller than the neutrino masses, and the differences can be safely neglected. If $\Delta m$ is small, the ordering becomes important. Figure adapted from \cite{King:2013}. 
	}
	\label{fig:hierarchy}
\end{figure}

The current knowledge of the neutrino mass distribution is summarised in \figref{hierarchy} for the three normal/active neutrinos ($\nu_e,\nu_\mu, \nu_\tau$) \cite{Fukuda:1998,Beringer:2012,King:2013}. If the value of $\Delta m$  (the mass of the lightest neutrino) is large, the mass differences are much smaller than the neutrino masses, and it is reasonable to assume the neutrinos have identical masses. We often refer to this as degenerate neutrinos and denote the scenario by \Mnuthree{} in the forthcoming analysis.

If $\Delta m$ is close to zero, the hierarchy will play a significant role. For the normal hierarchy there will be one neutrino with a mass close to the largest mass difference and two almost massless neutrinos. We call this model with one massive and two massless neutrinos \Mnutwo. For the inverted hierarchy there will instead be one massless and two massive species which we denote \Mnutwoone. 

For all of the above scenarios we keep the effective number of neutrinos, \Neff{}, fixed at $3.046$. However, Planck allows for extra radiation density at early times that can be parametrized as an increase in \Neff. We have varied \Neff{} for the \Mnuthree{} and \Mnutwo{} cases allowing for extra massless species (or any other dark radiation effect). These scenarios are called \Mnuthree+\Neff{} and \Mnutwo+\Neff{}.

Short baseline oscillation experiments have hinted at the existence of one or more sterile neutrino species with masses of the order of $1\eV$ \cite{Kopp:2011,Mention:2011, Huber:2011,Giunti:2011}. Even though such large masses are ruled out by structure formation if the neutrinos are thermalised \citep{Reid:2010,Thomas:2010,Riemer-Sorensen:2012,  dePutter:2012, Zhao:2013, Riemer-Sorensen:2013}, those constraints can be circumvented by non-standard physics mechanisms \cite[][]{Hannestad:2012,Steigman:2013,Hannestad:2013}. We have analysed one such short baseline-inspire scenario called \Mnufour{}. \Mnufour{} is parametrized as one massive specie with $m_3 = 0.06 \eV$ plus two massless neutrinos and one additional massive sterile neutrino for which we vary the mass \cite[similar to][]{Battye:2013,Wyman:2013}. \Neff{} can take any value, i.e.\ the sterile neutrino is not required to decouple at the same time as the active neutrinos. An earlier decoupling will lead to $\Delta N_\mathrm{eff}<1$ while later decoupling will lead to $\Delta N_\mathrm{eff}>1$.

\section{Data and method} \label{sec:method}

\subsection{Data}
The CMB forms the basis of all precision cosmological parameter analyses, which we combine with other probes. In detail, we use the following data sets:

\noindent \textbf{Planck:} The CMB as observed by Planck from the 1-year data release\footnote{\url{pla.esac.esa.int/pla/aio/planckProducts.html}} \cite{PlanckXVI:2013}. We use the low-$\ell$ and high-$\ell$ CMB temperature power spectrum data from Planck with the low-$\ell$ WMAP polarisation data (\textit{Planck+WP} in \cite{PlanckXVI:2013}). We marginalise over the nuisance parameters that model the unresolved foregrounds with wide priors, as described in \cite{PlanckXV:2013}. We do not include the Planck lensing data because they deteriorate the fit as described in \cite{PlanckXVI:2013}, implying some tension between the data sets, which will hopefully be resolved in future data releases. 

\noindent \textbf{BAO:} 
Both the matter power spectra and BAO are measured from the distribution of galaxies in galaxy-redshift surveys, and therefore one must be careful not to double-count the information. Thanks to the dedicated work of several survey teams we can choose from multiple data sets, and only use either the power spectrum or the BAO from any single survey. For the BAO scale we use the measurements from the Six Degree Field Galaxy Survey (6dFGS, $r_s/D_V(z=0.106) = 0.336\pm0.015$) \cite{Beutler:2011}, the reconstructed value from Sloan Digital Sky Survey (SDSS) Luminous Red Galaxies ($r_s/D_V(z=0.35) = 0.1126\pm0.0022$) \cite{Padmanabhan:2012}, and from the Baryon Oscillation Spectroscopic Survey (BOSS, $r_s/D_V(z=0.57) = 0.0732\pm0.0012$) \cite{Anderson:2012}. 

\noindent \textbf{WiggleZ:} For the full power spectrum information, we use the WiggleZ Dark Energy Survey\footnote{\url{smp.uq.edu.au/wigglez-data}} power spectrum \cite{Parkinson:2012} measured from spectroscopic redshifts of 170,352 blue emission line galaxies with $z<1$ in a volume of ~1 Gpc$^3$ \cite{Drinkwater:2010}, and covariance matrices computed as in \cite{Blake:2010}. The main systematic uncertainty is the modelling of the non-linear matter power spectrum and the galaxy bias. We restrict the analysis to $k<0.2$\hMpc{} and marginalise over a linear galaxy bias for each of the four redshift bins in the survey. 

\noindent \textbf{HST:} 
We also investigate the addition of a Gaussian prior of $H_0 = 73.8\pm2.4\, \km \, \s^{-1} \, \Mpc^{-1}$ on the Hubble parameter value today obtained from distance-ladder measurements \cite{Riess:2011}. Based on re-calibration of the cepheids Ref. \cite{Freedman:2012} found $H_0 = 74.3\pm2.1\,\mathrm{km\,s^{-1}\,Mpc^{-1}}$, and a different analysis by Ref. \cite{Riess:2011} found $H_0=74.3\pm2.1\, \km \, \s^{-1} \, \Mpc^{-1}$, which was subsequently lowered to $72.5\pm2.5\, \km \, \s^{-1} \, \Mpc^{-1}$ \cite{Efstathiou:2013} when the maser distances were re-calibrated \cite{Humphreys:2013}. Although slightly deviating, all the values remains consistent with the one adopted here.

\subsection{Parameter sampling}
We sample the parameter space defined in \secref{models} using the publicly available Markov Chain Monte Carlo (MCMC) sampler MontePython\footnote{\url{montepython.net}} \cite{Audren:2012} with the power spectra generated by CLASS \cite{Blas:2011}. The Planck likelihoods are calculated by the code provided with the Planck Legacy Archive\footnote{\url{pla.esac.esa.int/pla/aio/planckProducts.html}}. The WiggleZ likelihood is calculated as described in \cite{Parkinson:2012} but conservatively excluding the most non-linear part of the power spectrum by cutting at $k_\mathrm{max}=0.2$\hMpc (see \secref{range}). 

For a few scenarios we compared the MontePython samples to those of the publicly available CosmoMC \footnote{\url{http://cosmologist.info/cosmomc}} \cite{Lewis:2002} with the power spectrum generator CAMB \footnote{\url{http://camb.info}}. The results are very similar.

For random Gaussian data the $\chi^2$ per degree of freedom can be used to quantify the agreement between independent data sets. However, the Planck data likelihood is not Gaussian, and instead we compare the relative probability of the combined data to Planck alone
\begin{equation} \label{eqn:deltachi}
\Delta \chi^2/\Delta \mathrm{dof} \equiv 2\frac{\log\mathcal{L}_\mathrm{comb}-\log\mathcal{L}_\mathrm{Planck}}{\mathrm{dof}_\mathrm{comb}-\mathrm{dof}_\mathrm{Planck}}
\end{equation}
for the parameter likelihoods, $\mathcal{L}$, of a given model. We interpret this as a relative probability between Planck only and Planck+extra. If the increase in $\chi^2$ per extra degree of freedom is larger than 1, the relative probability of the two data sets is small (assuming they have been drawn from the same distribution), which implies a tension between the datasets. Such difference can originate from systematics in the data, inadequate modelling of the data, or an incorrect cosmological model. If $\Delta \chi^2/\Delta \mathrm{dof} \lesssim1$ the data sets are in statistical agreement.

\subsection{Priors}
We apply uniform probability priors on all parameters with a minimum of hard limits (given in \tabref{priors}). The limits that could be explored by the MCMC exploration were either set to be unbound in MontePython, or chosen to be very much wider than any expected posterior width in CosmoMC. All non-cosmological parameters introduced in the data likelihood codes are marginalised over. In particular we find that for neutrino masses close to the lower limit, the quoted value is very sensitive to the use of lower prior, and the literature is inconsistent on this point \cite[e.g.][]{PlanckXVI:2013,Hamann:2010,Parkinson:2012,Giusarma:2013,Feeney:2013,Reid:2010,Thomas:2010,Riemer-Sorensen:2012,dePutter:2012,Riemer-Sorensen:2013,Battye:2013,Wyman:2013,Hou:2012,Wang:2012,Joudaki:2013,Giusarma:2011}. Consequently in \tabref{likelihoods}, we quote the limits obtained with and without the lower prior. 

\begin{table}
\squeezetable
\begin{tabular}{l| l| l}
 Parameter 								& Starting value	& Prior range \\ \hline
$\Omega_\mathrm{b}h^2$					& 0.02207			& None $\rightarrow$ None		 \\
$\Omega_\mathrm{cdm}h^2$					& 0.1198			& None $\rightarrow$ None		 \\
$H_0$ [$\mathrm{km}\, \mathrm{s}^{-1}\, \Mpc ^{-1}$] & 67.3			& None $\rightarrow$ None		 \\
$A_s$ [$10^{-9}$]							& 2.2177			& 0 $\rightarrow$ None			 \\
$n_s$ 									& 0.9585			& 0 $\rightarrow$ None			 \\
$\tau$ 								& 0.091			& 0 $\rightarrow$ None			 \\
\Mnu{} [$\eV$] 								& 0.3				& 0.00 or 0.04 $\rightarrow$ None\\
\Neff{}	 								& 3.046			& Fixed or 0 $\rightarrow$ 7 \\
\end{tabular}
\caption{The parameters uniform probability priors for the MCMC sampling. In MontePython the prior edges were set to be unbound unless otherwise specified. The parameters are: baryon density ($\Omega_\mathrm{b}h^2$), dark matter density ($\Omega_\mathrm{cdm}h^2$), Hubble parameter ($H_0$), optical depth to reionisation ($\tau$), amplitude of the primordial density fluctuations ($A_s$), power spectrum index ($n_s$), sum of neutrino masses ($\sum m_\nu = N_\nu m_\nu$), effective number of neutrinos (\Neff{}).}
\label{tab:priors}
\end{table}

\subsection{Power spectrum range} \label{sec:range}
\begin{figure}
	\centering
	\includegraphics[width=0.49\textwidth]{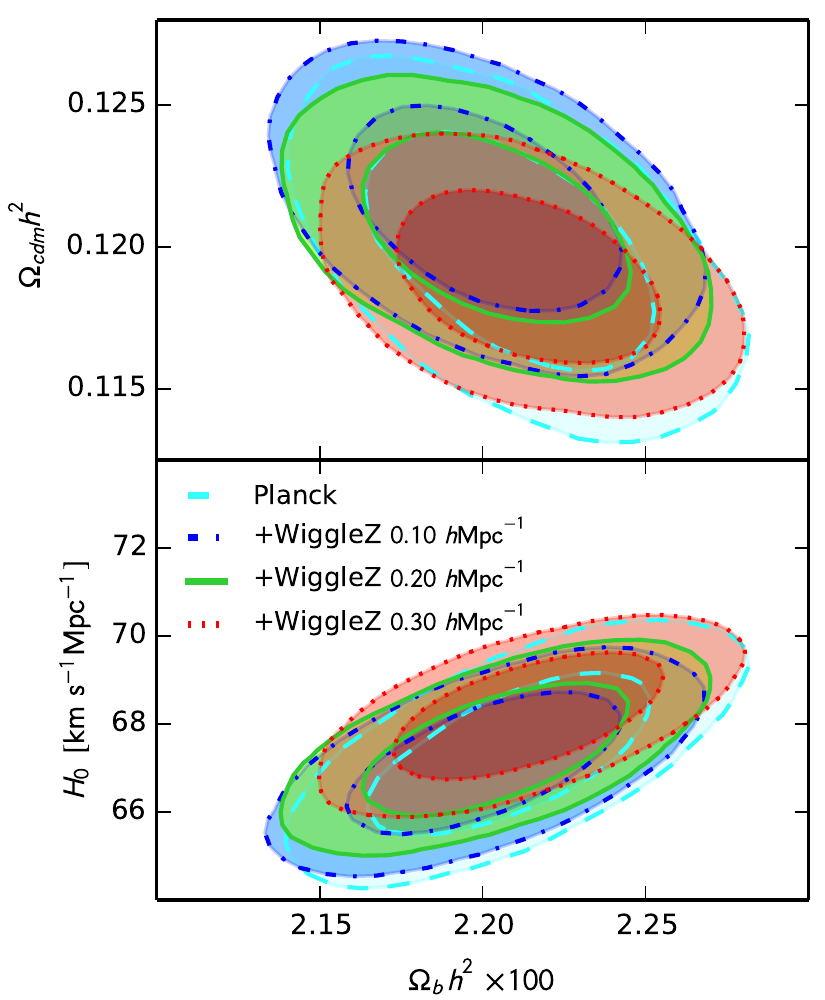}
	\caption{\LCDM{} fitted to Planck+WiggleZ as a function of \kmax{}. There is an excellent agreement between Planck and Planck+WiggleZ for all values of \kmax.}
	\label{fig:LCDM}
\end{figure}

Modelling the power spectrum on small scales where the linear theory for structure formation breaks down, is notoriously difficult.
To determine which \kmax\ cut-off provides the most robust constraints we analysed the Planck+WiggleZ data combination for \LCDM{} cosmology, varying $k_\mathrm{max}$ between $0.15$ \hMpc{} and $0.30$ \hMpc{}. The resulting parameter contours are shown in \figref{LCDM}.

There is an excellent agreement between Planck and Planck+WiggleZ for all values of \kmax. The agreement between fits with \kmax{} = 0.1 and 0.2\hMpc{} is good, but there is a small off-set for \kmax{} = 0.3\hMpc. The $\Delta \chi^2/\Delta \mathrm{dof} = [0.72, 0.81, 0.97]$ respectively, indicate a slight decrease in fit quality with \kmax. The decrease is worse for \kmax{} increasing from 0.2 to 0.3\hMpc{} than for 0.1 to 0.2\hMpc{} but all values are acceptable. 


For all further analyses we fix \kmax{} $=0.2$\hMpc{}. This throws out a lot of the power spectrum, which has measurements out to $k=0.5$\hMpc, but minimises the uncertainties in non-linear modelling. 

\begin{figure}
    \centering
    \includegraphics[angle=0,width=0.99\columnwidth]{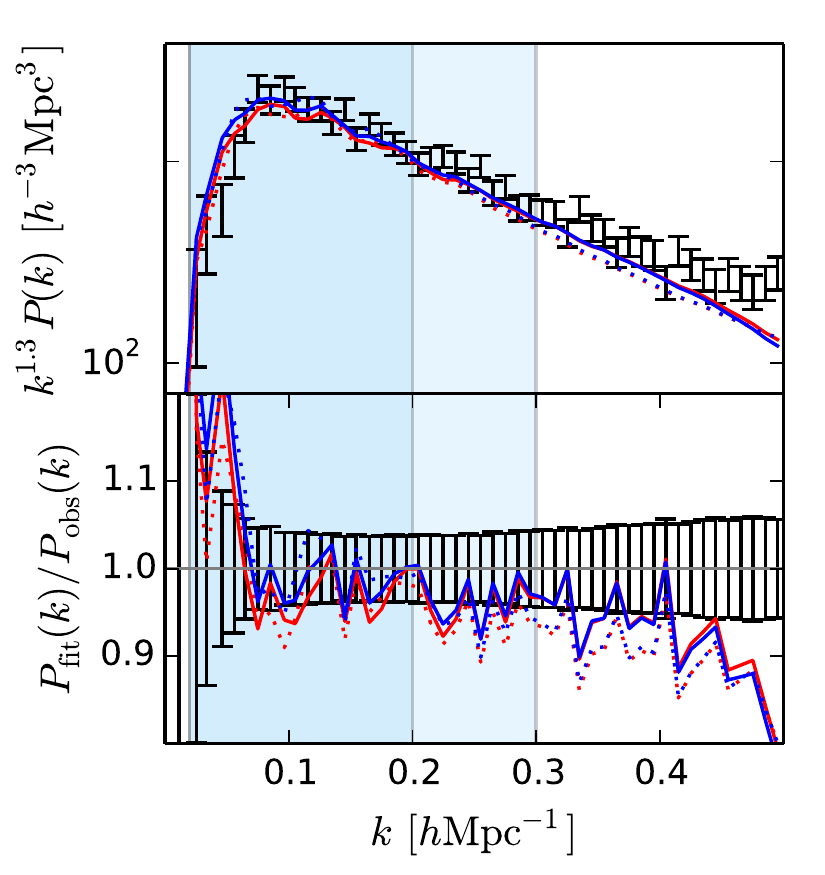}
    \caption{WiggleZ power spectrum averaged (for visualisation only) over the seven survey regions and four redshift bins (black bars) shown with the best fit \LCDM{} models for $k_\mathrm{max} = 0.2$\hMpc (red solid) and $k_\mathrm{max} = 0.3$\hMpc (blue solid) as well as the linear CLASS models for the same parameters (dotted, same colours). In the lower panel the models are compared after normalisation by the data values. 
    }
	\label{fig:powerspec}
\end{figure}

The best fit models of fits to Planck+WiggleZ to $k_\mathrm{max}=0.2$\hMpc{} and $0.3$\hMpc{} are shown in \figref{powerspec}. For $k<0.2$\hMpc{} the observed power spectrum fluctuates around both models, but for 0.2\hMpc$<k<0.3$\hMpc{} the model undershoots the data even when the range is included in the fit.

\subsection{Uncertainties of upper limits}
To check whether the differences between the models are real and not due to statistical sampling, we determine the uncertainty on the upper limit. The variance of the variance of a sample is given by\footnote{\url{http://mathworld.wolfram.com/SampleVarianceDistribution.html}}
\begin{equation}\label{eq:variance}
\mathrm{Var}(\sigma^2) = \frac{1}{n}\left(\mu_4 - \frac{n-3}{n-1}\sigma^4 \right) \, ,
\end{equation}
where $n$ is the independent sample size), $\sigma$ is the sample variance, and $\mu_4$ is the central fourth momentum of the underlying distribution (the kurtosis). For $n$ we use the number of independent lines in the MCMC chains as estimate provided by `GetDist' \cite{Raftery:1992}.
Since we quote $2\sigma$ (95\% confidence level) limits, we multiply by $2$,

\begin{eqnarray}\label{eqn:error}
\Delta \sum m_\nu (\mathrm{95\%}) =& \\ \nonumber
&2\sqrt{\frac{1}{n}\left(\mu_4(\sum m_\nu) - \frac{n-3}{n-1}\sigma(\sum m_\nu)^4 \right)} \, ,
\end{eqnarray}

The uncertainties on the \Mnu{} 95\% confidence limits are quoted in \tabref{likelihoods}. In most cases the difference between the models ($\sim0.02\eV$) are larger than the uncertainties ($\lsim 0.01\eV$). Consequently the differences cannot be attributed sampling effects alone.

\section{Results and discussion} \label{sec:results}
We list the fitted models and their best fit likelihoods in \tabref{likelihoods}, as well as $\Delta \chi^2/\Delta \mathrm{dof}$ and neutrino mass constraints with and without the low prior.

\begin{table*}
    \begin{tabular}{ l || c | c | c | c || c |c}
							& \multicolumn{4}{| c ||}{With lower prior of $\sum m_\nu > 0.04 \eV$} & \multicolumn{2}{c}{No lower prior}  \\ \hline
Data combination     				&-$\log \mathcal{L}$ &$\Delta \chi^2/\Delta \mathrm{dof}$ & \Mnu (95\% CL) [eV] 	& $\Delta$ \Mnu [eV] &-$\log \mathcal{L}$ 	& \Mnu (95\% CL) [eV]	\\ \hline
\hline
\multicolumn{7}{l}{\Mnuthree} \\ \hline
Planck\footnotemark[1] 		& 4902.6			& --- 			& 0.98		& 0.006			& 4902.6 		& 1.10 \\
Planck+BAO\footnotemark[1]	& 4903.0			& 0.23		& 0.35		& 0.006			& 4904.2		& 0.27\\
Planck+WiggleZ			& 5129.5			& 0.82		& 0.39		& 0.008			& 5129.6		& 0.35\\
{\bf Planck+BAO+WiggleZ}	& {\bf 5130.4}		&{\bf 0.81}		& {\bf 0.25}	& {\bf 0.008}		& {\bf 5130.8}	& {\bf 0.18}\\	
Planck+BAO+HST+WiggleZ	& 5134.0			& 0.82		& 0.19		& 0.020			& 5132.9		& 0.13\footnotemark[2]\\ 
\hline
\multicolumn{7}{l}{\Mnutwoone} \\ \hline
{\bf Planck+BAO+WiggleZ}	& {\bf 5130.8}		& ---			& {\bf 0.22}	& {\bf 0.015}		& {\bf 5130.5}	& {\bf 0.16}\\			
Planck+BAO+HST+WiggleZ	& 5134.0			& ---			& 0.17		& 0.009			& 5133.6  		& 0.13\footnotemark[2]\\ 
\hline
\multicolumn{7}{l}{\Mnutwo} \\ \hline
Planck\footnotemark[1]		& 4902.9			& ---			& 0.72		& 0.007			& 4902.4 	 	&0.73 \\ 
Planck+BAO				& 4903.4			& 0.39		& 0.30		& 0.010			& 4903.1		& 0.28\\
Planck+WiggleZ			& 5129.4			& 0.82		& 0.35		& 0.008			& 5129.4		& 0.18 \\
{\bf Planck+BAO+WiggleZ}	& {\bf 5130.2}		& {\bf 0.81}	& {\bf 0.21}	& {\bf 0.010}		& {\bf 5129.8}	& {\bf 0.16} \\	
Planck+BAO+HST+WiggleZ 	& 5133.4			& 0.82		& 0.17		& 0.009			& 5133.2		& 0.12\footnotemark[2]\\
\hline
\multicolumn{7}{l}{\Mnufour} \\ \hline
{\bf Planck+BAO+WiggleZ}	& ---				& --- 			& ---			& ---				&{\bf 5130.9}	& {\bf 1.51}\footnotemark[3]	\\	
\hline
\multicolumn{7}{l}{\Mnuthree+\Neff}  \\ \hline
{\bf Planck+BAO+WiggleZ}	& {\bf 5130.6}		& ---			& {\bf 0.37}	& {\bf 0.012} 		& ---			& ---	\\
 Planck+BAO+HST+WiggleZ	& 5131.7			& ---			& 0.41		& 0.014			& 5131.7		& 0.40		\\
\hline
\multicolumn{7}{l}{\Mnutwo+\Neff} \\ \hline
{\bf Planck+BAO+WiggleZ}	& {\bf 5130.9}		& ---			& {\bf 0.29}	& {\bf 0.014}		& ---			& ---  \\	

\end{tabular}
\footnotetext[1]{Results from CosmoMC} 
\footnotetext[2]{The inclusion of the HST prior may artificially enhance the constraint due to tensions between the data sets. In the \Mnutwo{} case $\Delta \chi^2/\Delta \mathrm{dof} =5.83$ for Planck+HST compared to 0.23 and 0.82 for Planck+BAO and Planck+WiggleZ, respectively. The values for \Mnuthree{} are very similar.}
\footnotetext[3]{Mass of the sterile species for which we set no lower prior}

    \caption{The best fit likelihood values and neutrino mass constraints for different assumptions about the hierarchy. We quantify the change in best fit likelihood when adding data to Planck alone by \eqnref{deltachi}. The additional degrees of freedom are: $\mathrm{dof}_\mathrm{WiggleZ} = 556$, $\mathrm{dof}_\mathrm{BAO} = 3$, $\mathrm{dof}_\mathrm{H0} = 1$. The sampling uncertainty, $\Delta$\Mnu{}, is determined by \eqnref{error}. In most cases it is smaller than the difference between the models. Notice how the \Mnu{} constraints tighten with the exclusion of the lower prior.
    }
    \label{tab:likelihoods}
\end{table*}

\subsection{Results: \Mnuthree}
\begin{figure*}
	\centering
	\includegraphics[width=0.49\textwidth]{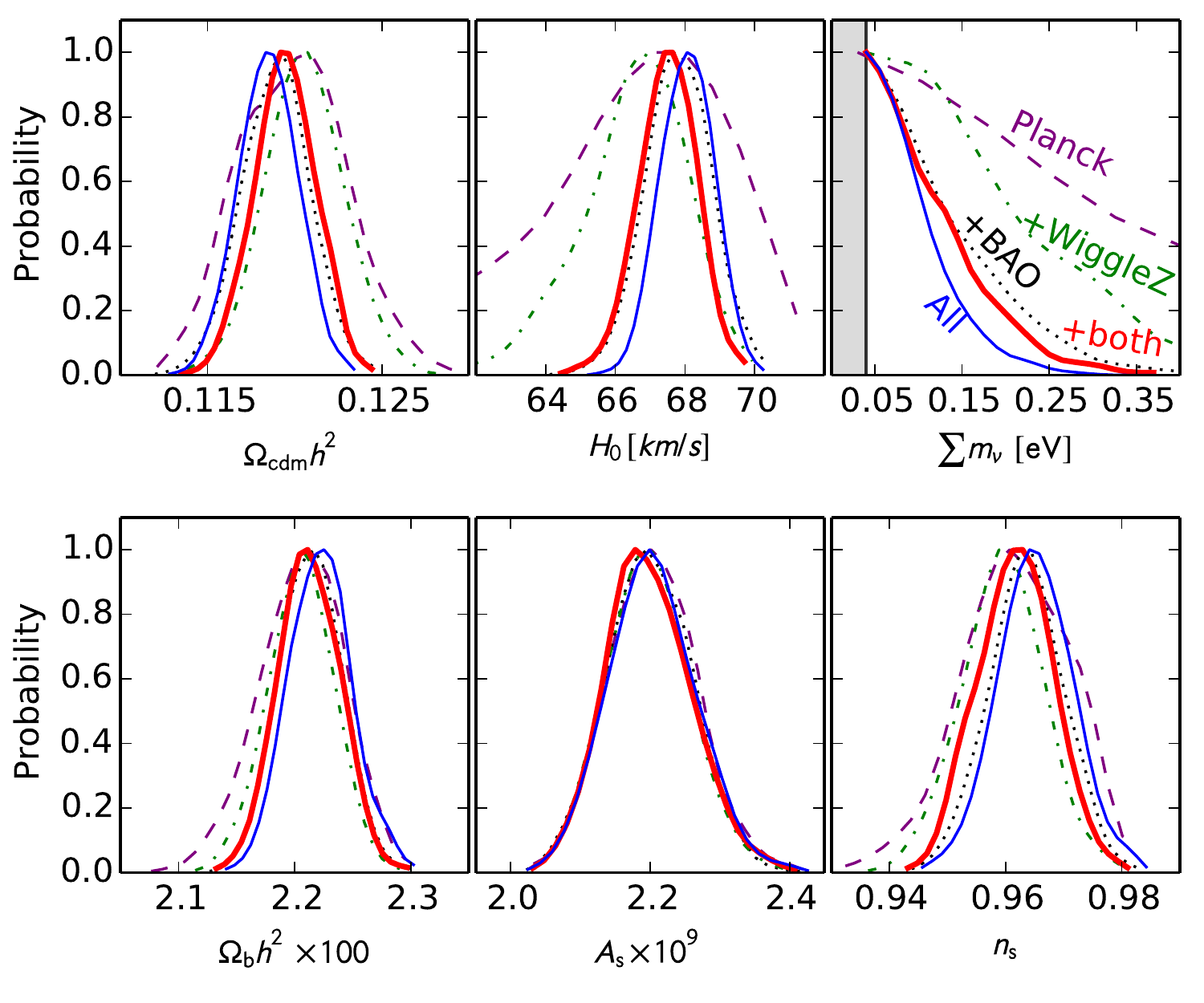}
	\includegraphics[width=0.49\textwidth]{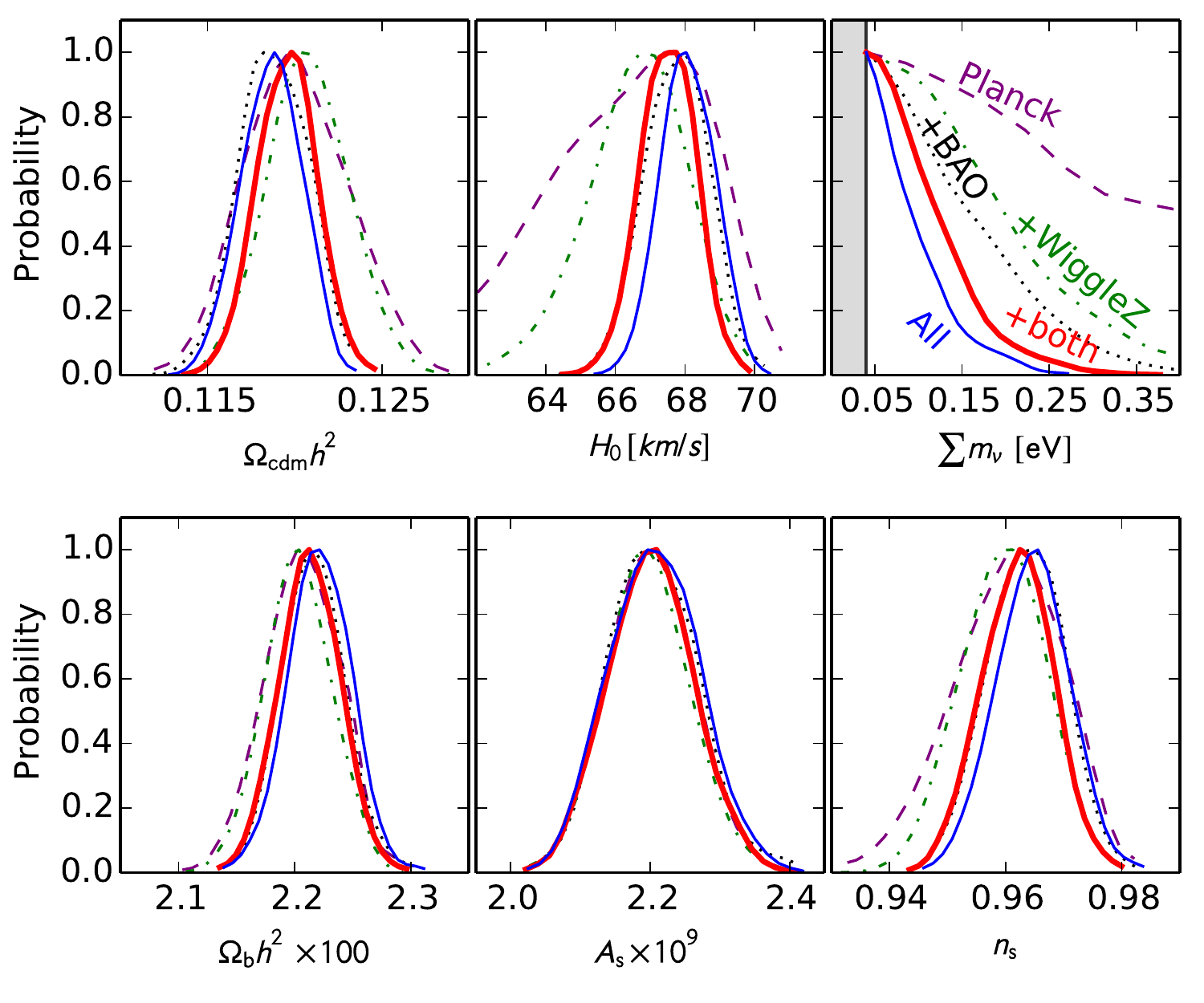}
	\caption{One-dimensional parameter likelihoods for fitting \Mnuthree{} (left) and \Mnutwo{} (right) to various data combinations: Planck (dashed purple), Planck+BAO (dotted black), Planck+WiggleZ (dot-dashed green), Planck+BAO+WiggleZ (thick solid red), Planck+BAO+HST+WiggleZ (thin solid blue). The main effect of adding other observations to Planck is a tightening of the constraints on $\Omega_\mathrm{cdm}$, $H_0$ and $\sum m_\nu$ (top row). The improvement of adding WiggleZ is more significant for \Mnutwo{} than for \Mnuthree{} indicating that the fit is sensitive to the power spectrum shape.}
	\label{fig:1dcont}
\end{figure*}

The left panel of \figref{1dcont} shows the one-dimensional parameter likelihoods for fitting \Mnuthree{} to various data combinations. The major differences occur for $\Omega_\mathrm{cdm}$, $H_0$ and $\sum m_\nu$ (top row). For $\Omega_\mathrm{cdm}$ and $H_0$ the constraints tighten relative to Planck alone. For \Mnu{} Planck+WiggleZ is better than Planck but worse than Planck+BAO. Adding WiggleZ to Planck+BAO only tightens the constraint slightly, but more importantly it does not introduce any tension like the one seen for other low redshift probes such as cluster counts and lensing data \cite{PlanckXVI:2013, Battye:2013, Wyman:2013}.

The Planck collaboration pointed out a tension between the Planck+BAO and local $H_0$ measurements \cite{PlanckXVI:2013}. This tension remains with the addition of WiggleZ and the obtained upper limit on \mnu{} may be artificially enhanced. 

If we disregard the information from particle physics and set the lower prior to zero, there is no sign of a preferred non-zero mass. However, the upper limit changes significantly from $0.25\eV$ to $0.18\eV$ for Planck+BAO+WiggleZ, and all the way down to $\sum m_\nu < 0.13\eV$ for Planck+BAO+WiggleZ+HST. The probabilities are very similar to those without a lower prior, but the 95\% confidence upper limit shifts downwards due to the area between 0 and 0.04 eV.

\subsection{Results: \Mnutwo}
\Mnutwo{} is the standard model neutrino scenario that differs most from \Mnuthree{}, since all the neutrino mass is in one specie rather than split over three.
The right panel of \figref{1dcont} shows the one-dimensional parameter probabilities of fitting \Mnutwo{} to various data combinations. Qualitatively the effect of WiggleZ is similar to the \Mnuthree{} case but more pronounced. The Planck+WiggleZ constraint on \Mnu{} is almost as good as the Planck+BAO constraint. Adding WiggleZ to the former significantly improves the constraint to $\sum m_\nu < 0.21 \eV$. The fact that WiggleZ performs differently for \Mnutwo{} and \Mnuthree{} indicates a sensitivity to the power spectrum shape. Three degenerate neutrinos will have a smaller effect smeared over a larger range of scales than one neutrino carrying the entire mass. At this stage we do not strongly constrain the hierarchy, 
as the  \Mnutwo{} scenario is only valid for $[\Delta m_{21} \approx 0.009 \eV] <<  [\Delta m_{32} \approx 0.05 \eV] \approx [\sum m_\nu]$, where one can safely model the neutrinos as one massive and two massless species (normal hierarchy model).  However, currently our upper limit $\sum m_\nu  \lsim 0.2 \eV$ is significantly higher than largest mass difference ($\Delta m_{32}$). 
Nevertheless, the fact that we are now seeing differences in constraints due to the different hierarchies reveals potential of near-future galaxy surveys.

\subsection{Results: \Mnutwoone}
\begin{figure}
	\centering
	\includegraphics[width=0.49\textwidth]{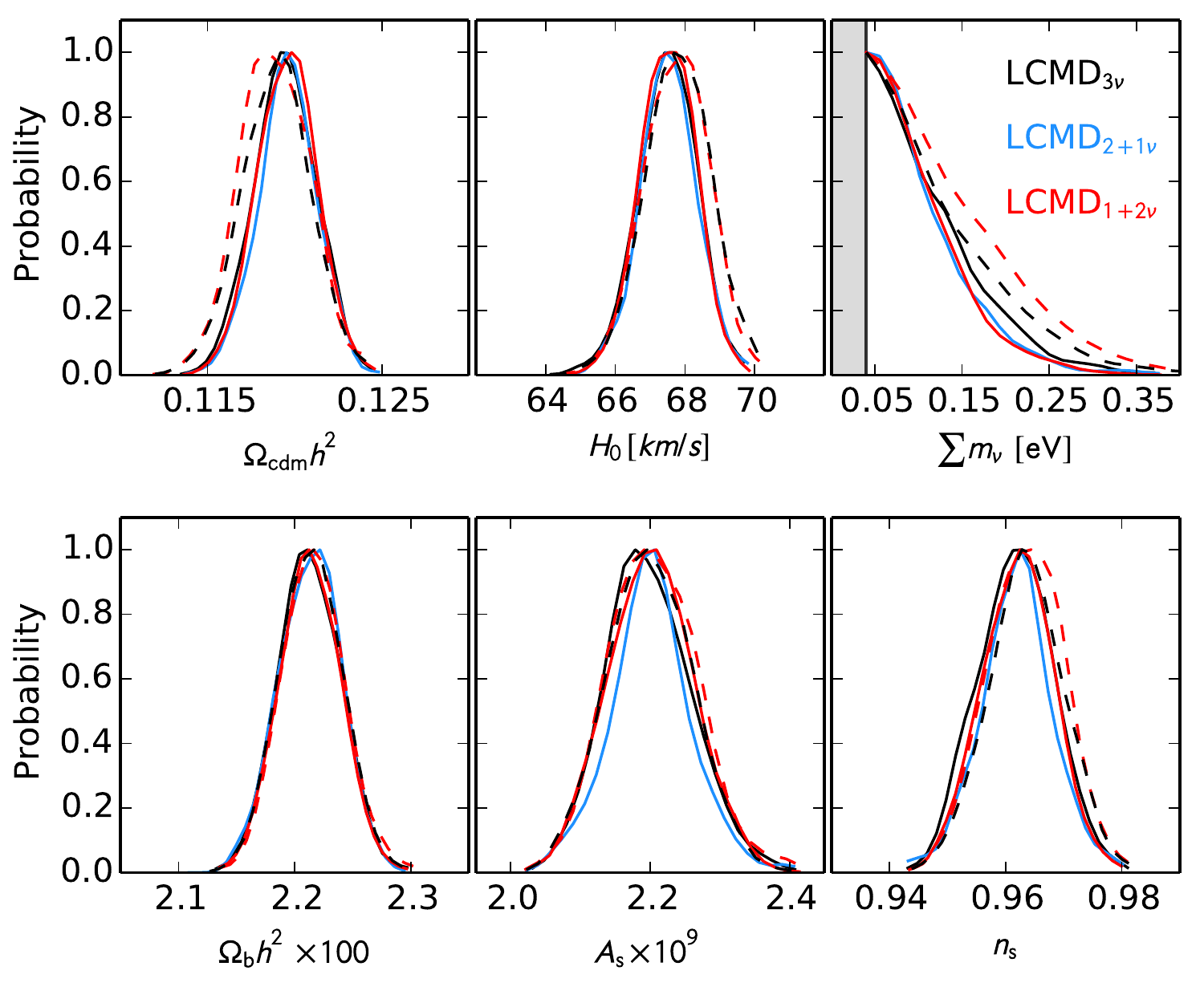}
	\caption{One-dimensional parameter probabilities comparing \Mnutwo{} (red), \Mnutwoone{} (blue), and \Mnuthree{} (black) fits to BAO+Planck+WiggleZ (solid) and Planck+BAO (dashed). None of the preferred parameter shifts significantly between the different scenarios, only the \Mnu{} limit changes.}
	\label{fig:compare}
\end{figure}

\figref{compare} shows the one-dimensional parameter probabilities comparing \Mnutwo{}, \Mnutwoone{}, and \Mnuthree{} fits to Planck+BAO+WiggleZ. There is no apparent change in the preferred parameter values between the models. The only significant difference is the tightness of the \Mnu{} constraints. For \Mnuthree{} Planck+BAO is slightly stronger than Planck+BAO+WiggleZ, whereas the opposite is true for \Mnutwo. Somewhat surprisingly \Mnutwoone{} is almost identical to \Mnutwo{} and does not fall in the middle between \Mnutwo{} and \Mnuthree{}.

\subsection{Results: \Mnufour}
Refs. \cite{Battye:2013,Wyman:2013} found that the tension between Planck and lensing or clusters can be relieved by the addition of a massive sterile neutrino. We investigated this scenario and as it provides a fit that is equally good fit as \Mnuthree{}, the conclusion is that BAO+Planck+WiggleZ still allows the existence of such a massive sterile neutrino, but does not add to the evidence of its possible existence. 

\subsection{Results: \Mnuthree+\Neff{} and \Mnutwo+\Neff}
Before Planck, the addition of the effective number of relativistic degrees of freedom as a free parameter led to a significant weakening of the neutrino mass constraints \cite{Hamann:2010,Hou:2012,Riemer-Sorensen:2013,Wang:2012,Joudaki:2013,Giusarma:2013a}. Now, with the inclusion of higher multipoles, the Planck data suffers only mildly from this effect, and therefore it is less important to simultaneously fit for \Neff\ when fitting for $\sum m_\nu$. Nevertheless, the Planck results did leave space for extra species, and it remains interesting to fit for \Neff. Doing so, we find $N_\mathrm{eff} =3.28^{+0.42}_{-0.26}$ (95\% confidence), and a weaker upper limit of \mnu$<0.37\eV$ for Planck+BAO+WiggleZ (with the lower prior). Although the Planck results alone gave no strong support for extra species, they still sat at $N_\mathrm{eff} = 3.36^{+0.68}_{-0.64}$ for Planck alone\footnote{including the high-$\ell$ data from South Pole Telescope \cite{Story:2013,Reichardt:2013} and Atacama Cosmology Telescope \cite{Das:2013}} or $N_\mathrm{eff}=3.52^{+0.48}_{-0.45}$ when combined with BAO and $H_0$, approximately $2\sigma$ above the standard $N_\mathrm{eff}=3.046$. 

Combining with large scale structure measurements, as we have done here, now prefers extra species at the $1\sigma$ level ($3.28^{+0.42}_{-0.26}$), and $2\sigma$ when including HST (\Neff{}=$3.40^{+0.44}_{-0.35}$, both values are 95\% confidence levels). The preferred value of \Neff{} is identical for \Mnuthree{} and \Mnutwo.

Allowing for extra neutrino species alleviates the tension between Planck+BAO and HST \cite[as also noted by][]{PlanckXVI:2013}, 
and also with the low redshift probes like galaxy cluster counts and gravitational lensing \cite{Battye:2013,Wyman:2013}. This remains true with the addition of WiggleZ, but at the cost of \Neff{} above the standard value. As mentioned in \cite{Audren:2012} the preference for high \Neff{} might simply originate in lack of understanding of late time physics.


\subsection{Non-linear scales}
On the quasi-linear scales up to \kmax{} = 0.2\hMpc{} the bias of the blue emission line galaxies in WiggleZ is linear to within 1\% \cite{Poole:inprep}. Adding a different shape dependent parametrisation will degrade the \Mnu{} constraints significantly. It is out of the scope of this paper to model additional non-linear effects, but we notice that for \Mnuthree{}, reducing the fitting range of WiggleZ to \kmax{} = 0.1\hMpc{} the constraint changes from $0.25\eV$ to $0.26\eV$ for the low prior fit to Planck+BAO+WiggleZ (compared to $\sum m_\nu <0.35$ for Planck+BAO alone).

\subsection{Measuring hierarchy}
\begin{figure*}
	\centering
	\includegraphics[trim = 0mm 0mm 0mm 70mm, clip, width=0.99\textwidth]{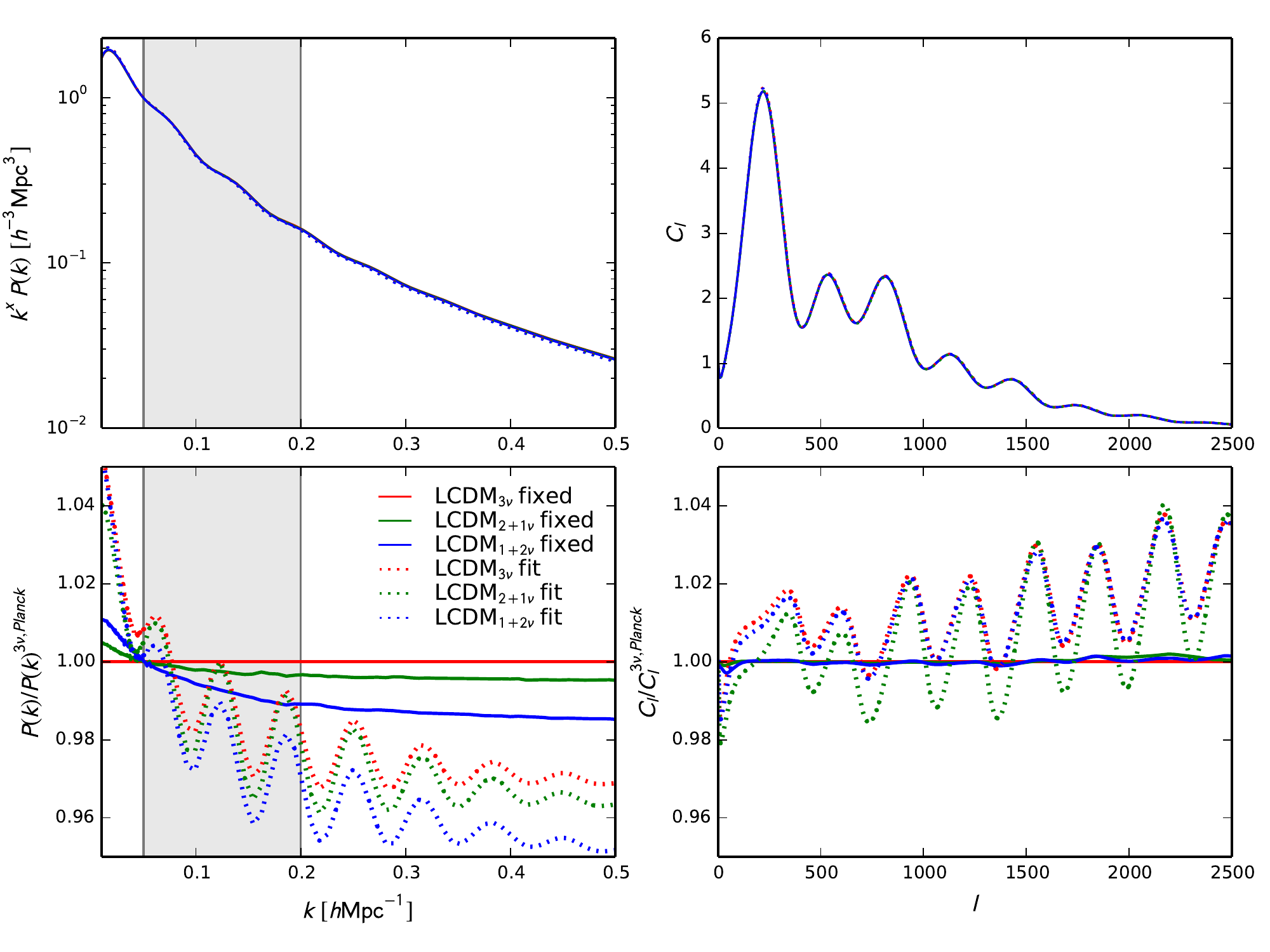}
	\caption{The ratio of power spectra for three different hierarchy scenarios relative to \Mnuthree{}. The left panel shows the matter power spectra, while the right is the CMB power spectra. The solid lines illustrate the magnitude of the hierarchy effect -- these models all have the same cosmological parameters (Planck best fit values and $\sum m_\nu = 0.15 \eV$), and differ only in the type of neutrino hierarchy assumed. The difference in the CMB power spectrum is negligible, but the matter power spectra differ by a few percent. The dotted lines show the best fit models for Planck+BAO+WiggleZ. The different hierarchies lead to best fit power spectra that are very similar, due to the degeneracy between the preferred values of $\Omega_\mathrm{cdm}$, $H_0$, and \Mnu{}.}
	\label{fig:Pk}
\end{figure*}

To investigate the possibility of measuring the hierarchy, we have compared the theoretical matter power spectra for the different scenarios to the uncertainty of the present day state of the art observations. \figref{Pk} shows the ratio of the matter and CMB power spectra relative to \Mnuthree. For a fixed cosmology (solid lines) the difference in the CMB power spectrum is negligible, but the matter power spectra differ by a few percent for $\sum m_\nu = 0.15 \eV$. The effect is mainly apparent on large scales, and can consequently be measured from the linear power spectrum alone. The dotted lines show the individual best fits to Planck+BAO+WiggleZ (also normalised to \Mnuthree). The degeneracies between neutrino mass and $\Omega_\mathrm{cdm}$ and $H_0$ lead to three very similar curves. It will be impossible to distinguish the hierarchies from the CMB alone, but the addition of large scale structure information can potentially distinguish between hierarchies based on linear scales alone. As inferred from the different neutrino mass limits obtained for the different scenarios, the combined analysis is already sensitive to the difference, but there is not enough difference in the likelihoods, yet, to determine the hierarchy.

\section{Summary and conclusions} \label{sec:summary}
We draw the following conclusions:

\begin{itemize}
\item There is good agreement between Planck and WiggleZ data, when using the value of \kmax{} = 0.2\hMpc{} for WiggleZ (\figref{LCDM}).
\item We have presented the strongest cosmological upper limit on the neutrino mass yet published, $\sum m_{\nu} < 0.18 \eV$ for a $\Lambda$CDM model with $\sum m_{\nu}$ as a free parameter.  
\item WiggleZ makes a larger difference for \Mnutwo{} than for \Mnuthree{}. This may indicate sensitivity to the power spectrum shape (\figref{compare}) as we would expect all the neutrino mass in one specie to suppress the power spectrum more than the case where it is equally distributed over three species (for the same total mass).
\item The uncertainties on the 95\% CL upper limits on \Mnu{} are smaller than the actual differences between the models, so the differences cannot be explained by sampling alone, but originate in the different models and priors.

\item There is no effect on the contours from the lower prior on \Mnu{} (\figref{compare}), but the 95\% CL limit changes (due to the area between 0 and 0.04 eV).
\end{itemize}

The improvement from adding WiggleZ to BAO+Planck and the sensitivity to the power spectrum shape bodes very well for potential constraints from future large scale structure surveys \cite{Giusarma:2011,Eisenstein:2011,Lahav:2010,Euclid}. Given the lower limit from particle physics, the allowable range for the sum of neutrino masses is $0.05\eV < \sum m_\nu < 0.25\eV$. In the inverted hierarchy (two heavy and one light neutrino) the neutrino oscillation results require $\sum m_\nu>0.1\eV$. If next generation of large scale structure surveys push the mass limit below $\sum m_\nu < 0.1\eV$, the inverted hierarchy can be excluded (under the assumption that \LCDM{} is the correct description of the universe).

The issue of high \Neff\ remains an open question. The combination of Planck+BAO+WiggleZ data prefers more than three neutrino species.

Neutrino mass constraints are important goals of current and future galaxy surveys \cite{Giusarma:2011} such as Baryon Oscillation Spectroscopic Survey \cite{Eisenstein:2011}, Dark Energy Survey \cite{Lahav:2010}, and Euclid \cite{Euclid}. Even stronger constraints on both $\sum m_\nu$ and \Neff\ would be achievable if we were able to use the whole observed matter power spectrum in the non-linear regime.  Currently we are not data-limited, but theory-limited in this area.  Improved theoretical models and simulations of the non-linear structure formation and redshift space distortions are crucial not only for future data sets, but also if we are to fully utilise the large scale structure data we already have in hand. 

\acknowledgements
We would like to thank Benjamin Audren for excellent support with MontePython, and Chris Blake for useful and constructive comments on the draft.  
TMD acknowledges the support of the Australian Research Council through a Future Fellowship award, FT100100595.  We also acknowledge the support of the ARC Centre of Excellence for All Sky Astrophysics, funded by grant CE110001020.

\bibliography{references} 

\end{document}